# *Effect of Niobium Doping on the Crystal Structure and Hydrogen Sorption Properties of TiFe: Combined Synchrotron X-ray Diffraction and Extended X-ray Absorption Fine Structure Study*

*Abhishek Banerjee [a], Stefano Deledda [b], Olena Zavorotynska [a\*]*

[a] Department of Mathematics and Physics, University of Stavanger, Stavanger P.O. Box 8600, NO-4036 Forus, Norway
[b] Department of for Hydrogen Technology, Institute for Energy Technology, P.O. Box 40, NO-2027 Kjeller, Norway

*olena.zavorotynska@uis.no

**Abstract**

TiFe alloys are attractive compounds for solid-state stationary hydrogen storage. They can absorb hydrogen gas reversibly at near ambient temperatures and practical pressures with high volumetric capacities surpassing that of cryogenically liquified $H_2$. The main drawback of TiFe-based storage systems is a costly activation procedure required due to the formation of oxide surface layer, which hinders hydrogen diffusion into the bulk. Doping the alloy with various additives is known to improve hydrogen diffusion softening the conditions of the activation procedure. Hydrogen sorption properties of the modified alloys have been the focus of most studies whereas less attention has been dedicated to the fundamental understanding of the effects of hydrogen sorption on the alloys' structure. The latter, however, is an important information in the knowledge-guided design of novel materials.

In this work, we investigated effects of Nb-doping on crystallographic structure of TiFe metal-alloy compounds and their hydrogen sorption properties. TiFe samples with two different Nb stoichiometries were synthesized using arc-melting (AM) and characterised with synchrotron powder X-ray diffraction (SR-PXRD) and extended X-ray absorption fine structure (EXAFS) analysis. Overall, $H_2$ absorption measurements (at 50 ± 2 ºC and 40 ± 2 bar), have shown that doping of TiFe with Nb can improve matrix activation and kinetics of hydrogen sorption without compromising the overall storage capacities. Refinement of SR-PXRD and EXAFS data showed significant Nb occupancy in secondary Ti phases, which improved the hydrogenation properties of the alloys.

**Keywords**
EXAFS, Nb additives, TiFe alloys, in situ synchrotron X-ray diffraction, hydrogen storage

**Graphical Abstract**

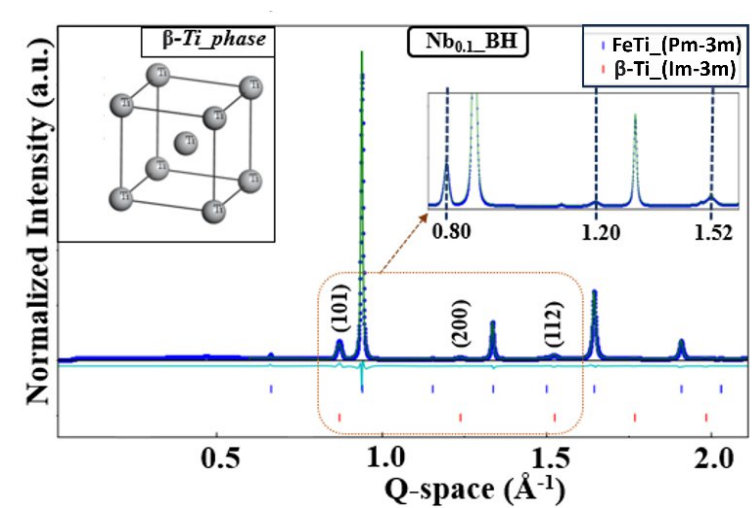
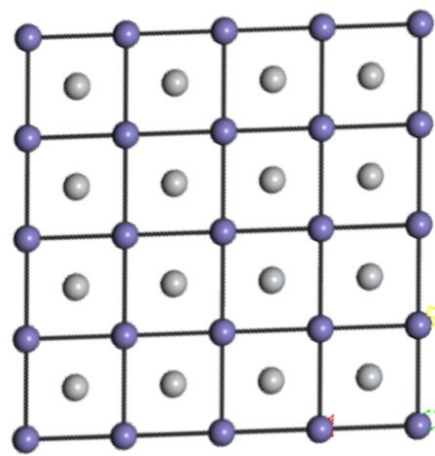
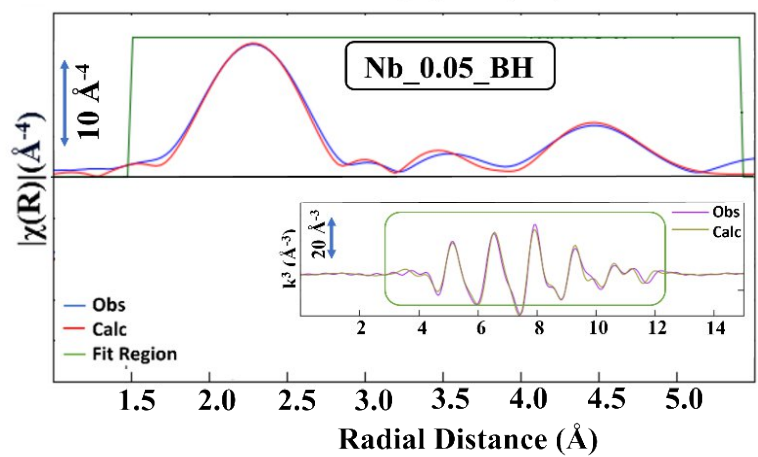
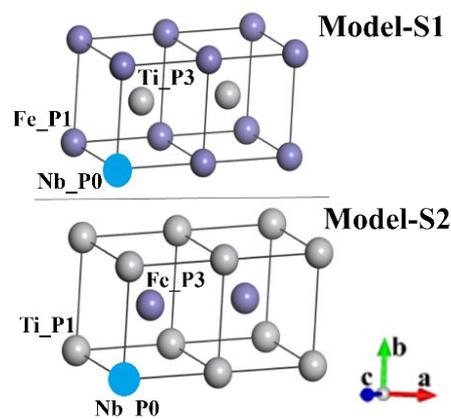

**Long-range and local structure of TiFe-Nb doped metal-alloys from SR-PXRD and EXAFS**

# 1. Introduction

TiFe intermetallic has been introduced as a practical storage medium several decades ago. It has the ability to absorb and release hydrogen at or near ambient temperatures (30-70 ºC) and low pressures (1-2 MPa), providing a major safety advantage **(1)**. TiFe also possesses other favourable characteristics such as non-toxicity, low cost, natural abundance in raw materials (Fe and Ti are the fourth and ninth most abundant metals on earth respectively), high volumetric hydrogen absorption capacity (0.096 $kg_{H2}$/L) and recyclability **(1-5)**. These properties make TiFe and its derivatives suitable for on-site stationary storage and maritime applications where the gravimetric hydrogen density is not critical **(6)**. Major disadvantages include susceptibility to surface oxidation that hinders hydrogen surface diffusion and requires extensive activation conditions (> 100 ºC and 10 MPa of $H_2$). Low gravimetric hydrogen capacity (< 1.9 wt%) limits its use as portable storage medium **(1-3, 7)**. Therefore, extensive research have been conducted on TiFe hydrides to understand its crystallographic properties **(8-13)**, hydriding / dehydriding mechanisms, and rate-determining steps in the sorption reactions **(14-16)**.

Equiatomic TiFe phase forms from molten liquid within narrow composition range (47.5 – 50.3 at%) at room temperature **(17)**. The intermetallic compound (IMC) has CsCl-cubic crystal structure (*Pm-3m*, No. 221) with lattice parameter 2.972 Å and Fe at (0, 0, 0) **(13, 18, 19)**. The unit cell hosts 12 tetrahedral and 6 octahedral interstices. Each tetrahedral site is surrounded by two Ti atoms and two Fe atoms forming the $Ti_2Fe_2$ configuration. Given the presence of two types of elements, two octahedral interstices exist: $Ti_4Fe_2$ and $Ti_2Fe_4$ **(20, 21)**. The larger atomic radius of Ti results in a larger interstice size for $Ti_4Fe_2$ (0.32 Å) compared to $Ti_2Fe_4$ (0.17 Å). $Ti_4Fe_2$ interstices are preferred adsorption sites for H as they offer a more negative hydrogen absorption energy **(7)**. Tetrahedral sites with a larger radius (0.38 Å) are usually filled up only at high pressures and high H content after all the octahedral sites have been occupied. Both sites participate in hydrogen diffusion **(22, 23)**.

Three TiFe hydride phases — namely, $α$-TiFeH$_{0.1}$, $β$-TiFeH$_{1.04}$, and $γ$-TiFeH$_{1.90}$ — are identified based on hydrogen concentration during isothermal P–C hysteresis **(1)**. The $α$-phase entails a solid solution with H filling $Ti_4Fe_2$ sites expanding the cubic TiFe lattice. Formation of $β$-phase causes lattice distortion, transforming the cubic lattice into an orthorhombic lattice (*P*222$_1$, No.17) with partial (90 %) filling of the octahedral interstices **(9, 10, 13)**. The $γ$-phase exhibit further lattice distortion accommodating more H atoms and is predicted to have an orthorhombic symmetry (space group *Cmmm*, No.65) with complete occupation of the octahedral $Ti_4Fe_2$ interstices and partially filled $Ti_2Fe_4$ sites **(10)**.

Oxide formation in TiFe intermetallic poses significant challenges for hydrogen sorption and thus storage applications. The presence of oxide layers hinder $H_2$ absorption and desorption kinetics leading to reduced storage capacity and cycling stability **(1, 24)**. This issue becomes more pronounced at elevated temperatures, where the oxide layer grows thicker, further impeding hydrogen diffusion **(2)**. Various methods, such as alloying **(25-27)**, nanostructuring **(28, 29)**, and inducing strain and cracks **(29, 30)**, have been employed to improve the activation kinetics of TiFe, which is crucial for practical implementation as a solid-state storage medium. Incorporation of different elements (for ex. Mg, Be, Zr, Cr, Hf, V, Co, Ta, Mo, Ni, Pd, Cu, Ce, La, Mm) as substitutional impurities in pristine TiFe alloys has shown an improvement of the activation, uptake kinetics and storage properties **(31)**. Alloying can induce effects like surface modification **(24, 32)**, enhance surface catalytic activity facilitating dissociation of $H_2$ molecules and improving the adsorption kinetics, stabilize crystallographic structure reducing probability of defects and vacancies which are known to accelerate oxidation processes (24, 33). Oxide layer thinning has been also observed with Mn dopants, which can modify the growth kinetics and limit oxidation extent (33).

To the best of our knowledge, very few studies of TiFe doped with Nb have been performed to date and a clear correlation of its crystallographic structure and hydrogen storage properties is missing. First-principle density-functional theory (DFT) study by Bakulin et al. (23) showed that dopants could potentially replace Ti ($r_a$ = 0.147 nm, $m_a$ = 47.87 u) in its respective lattice position increasing lattice size. Lattice expansion could improve H diffusion kinetics in TiFe matrix. In addition, experimental studies found that the doping resulted in the formation of secondary phases that can also be responsible for the improvement of the sorption properties (23). Cao et al. (34) showed that incorporating Sn and Nb in TiFe matrix induced formation of $\beta$-Ti phase and that the larger lattice mismatch between the $\beta$-Ti and TiFe phases in Nb-modified eutectic Ti–Fe alloys introduced strain at the interface. Nagai et al. (35) also observed the formation of precipitate phases and suggested that the interfaces between the FeTi matrix and the precipitates can act as active sites for the hydriding reaction and/or as entrance sites for hydrogen to diffuse into the alloy. This allowed to achieve room temperature activation of TiFe without prior heat treatment and shorten the initial hydriding incubation time. The precipitation of Ti-rich phases would indicate Nb substitution at Ti sites, which is most plausible given similar atomic radii. Berdonosova et al. (36) reported $H_2$ activation and sorption kinetics upon Nb doping up to 2 at%, showing the complete formation of $\gamma$-dihydride phase at ~7 MPa. They have also reported the doping concentration limit of 2 at% for Nb and Co required for formation of single phase TiFe-type compounds. Lattice expansion of the main $\beta$-TiFe phase was shown in the above-mentioned studies, suggesting Nb incorporation. Direct evidence of the location of Nb atoms and their local structure in the lattice and possibly in the related secondary phases, its correlation with Nb content and post $H_2$ absorption / desorption crystallographic structural changes have not been reported.

In this study, TiFe samples with varying Nb stoichiometries (0.05, 0.1) were synthesized using arc-melting (AM). The location of the Nb dopant in the CsCl-type crystal structure of TiFe before and after hydrogen adsorption was studied utilizing Rietveld refinement of synchrotron powder X-ray diffraction (SR-PXRD) data and local structure analysis by extended X-ray absorption fine structures (EXAFS). We have also studied the linear expansion coefficients of the pristine and doped matrices with in-situ synchrotron X-ray diffraction. Hydrogen sorption reaction kinetics was analysed using well-established diffusion models (37-40).

## 2. Experimental

### 2.1. Sample Preparation

The alloys were produced by arc-melting (AM) using an Edward Buhler MAM-1 arc melter. Ti (99.7 % metals basis, slug), Fe (99.9% metals basis, slug), and Nb (99.8% metals basis, slug) were purchased from Fisher Scientific AS and stored in a glove box (MBRAUN Inert gas System Company) under inert Ar atmosphere with $O_2$ and $H_2O$ levels less than 1 ppm. The melt chamber of the arc-melter was purified by melting a Ti-getter piece to reduce $O_2$ contamination in the chamber and then Ar flushed 3-4 times before sample preparation.

The slug pieces were cut in air using metal cutter prior to arc-melting. The melting was performed in a water-cooled copper crucible in high purity Ar atmosphere with parameter specifications (700 mbar, 60 amps). The samples were flipped over and re-melted four times to improve their chemical homogeneities. Four samples with stoichiometries: TiFe, TiFe$_{0.9}$, TiFe$_{0.9}$Nb$_x$ (x = 0.05, 0.1) were prepared. All samples showed less than 0.1% weight loss. The prepared ingots were pulverized with a table-top hammer and the residues were sieved through a 100 μm Al mesh. Particles less than 100 μm, were used for further XRD, EXAFS and hydrogen absorption studies. The sample abbreviations which will be used hereafter in this study are presented in Table 1.

**Table 1.** Sample list, abbreviations which are used in this study hereafter. "After H$_2$" stands for after an activation + 1 H$_2$ absorption / desorption cycle.

| Nominal Compositions (at%) | Sample Abbreviations |
|---|---|
| **TiFe**: *Before hydrogenation* | TiFe_BH |
| **TiFe**: *After H$_2$ act/abs/desorption* | TiFe_AH |
| **TiFe$_{0.9}$**: *Before hydrogenation* | TiFe$_{0.9}$_BH |
| **TiFe$_{0.9}$**: *After H$_2$ act/abs/desorption* | TiFe$_{0.9}$_AH |
| **TiFe$_{0.9}$Nb$_{0.05}$**: *Before hydrogenation* | Nb$_{0.05}$_BH |
| **TiFe$_{0.9}$Nb$_{0.05}$**: *After H$_2$ act/abs/desorption* | Nb$_{0.05}$_AH |
| **TiFe$_{0.9}$Nb$_{0.1}$**: *Before hydrogenation* | Nb$_{0.1}$_BH |
| **TiFe$_{0.9}$Nb$_{0.1}$**: *After H$_2$ act/abs/desorption* | Nb$_{0.1}$_AH |

## 2.2. SR-PXRD and SEM

The synchrotron powder X-ray diffraction (SR-PXRD) patterns were obtained at BM01 end station of the Swiss-Norwegian Beamlines (SNBL), at European Synchrotron Radiation Facility (ESRF) in Grenoble, France. The samples were loaded in 0.5 mm glass capillaries, and the patterns were collected with the Pilatus @ SNBL diffractometer (41) in Debye-Scherer geometry at room temperature. Data were collected in transmission mode with acquisition time of 10s, and 2D images were treated (masked for parasitic regions and integrated into 1D diffractograms) using BUBBLE (41). The distance between detector and sample was fixed at 0.139 m.

Experiments were performed in two sets: before and after H$_2$ activation followed by one cycle of absorption/desorption from the samples (details of H$_2$ exposure is provided in Section 2.4.), with wavelengths 0.6913 Å and 0.7120 Å, respectively. The wavelengths were calibrated using LaB$_6$ (NIST 660a standard). For in-situ measurements as function of temperature (300-500 K), the wavelength of 0.6913 Å was also used and the temperature was changed using a heat blower at 2K/min.

All diffraction patterns were refined using GSAS II software (42-44). The instrumental parameters contributing to peak broadening: Gaussian ($G_U$, $G_V$, $G_W$) and Lorentzian ($L_X$, $L_Y$, $L_Z$) terms were obtained from pattern fitting LaB$_6$ standard. The calculated instrumental parameter values for two experimental sets with different wavelengths are provided in Table S1.

Micro-structural images of samples were obtained using scanning electron microscopy (SEM; JEOL JSM-6610LV), operated at 10 keV and 30 mA at different magnifications to understand alloy morphology. The average composition of the alloys were investigated by energy dispersive spectroscopy (EDS) analysis at different sample positions.

**2.3. EXAFS measurements and data analysis**

Extended X-ray absorption fine structure spectroscopy (EXAFS) data on samples $Nb_{0.05}\_BH$, $Nb_{0.05}\_AH$, $Nb_{0.1}\_BH$ and $Nb_{0.1}\_AH$ were collected at Nb k-edge (~ 19 KeV) in transmission mode at BM31 end station of the SNBL-ESRF. The samples were measured as cellulose-diluted pellets which were produced using XAFSmass software **(45)**.

The white beam was monochromatized using an air-bearing liquid nitrogen Si (111) double crystal monochromator; harmonic rejection was performed by using two flat mirrors. EXAFS spectra were collected in transmission mode up to $k$-space (15 Å$^{-1}$). Ionization chambers measuring $I_0$ and $I_1$ were optimized to monitor intensities before and after assembling. For each sample, three consecutive EXAFS spectra were collected and averaged to minimize noise **(46)**. The EXAFS spectra were extracted and analysed using Demeter package **(47)**. The EXAFS function $\chi(k)$ was extracted using Athena code **(48)**. The averaged $k^3\chi(k)$ function was Fourier transformed in the interval 2.00–11.00 Å$^{-1}$. The fits were performed in $R$-space in the range 1.5–5.5 Å.

The phases and amplitudes for the single scattering (SS) and multiple scattering (MS) paths were calculated using FEFF6 code **(46)** within the Artemis software **(47, 48)**. TiFe, $\beta$-Ti and $\alpha$-Ti crystallographic information files from Materials Project **(49-54)** were used as inputs for the EXAFS fitting. It was assumed that Nb can occupy both Ti and Fe lattice sites in the crystal structure and Ti sites in formed $\alpha$ and $\beta$-Ti secondary phases. Thus two different TiFe supercells with large enough sizes (4*4*4) to uniformly distribute Nb stoichiometries (0.05, 0.1) were built using VESTA software **(55)**. Additional supercells, $\alpha$ and $\beta$-Ti were used to model Nb placement in the formed secondary α and β-Ti phases in the AH and BH samples, respectively. The calculated scattering paths contributing to intensities more than 10% to the overall model function, were considered further for fitting experimental data. Fittings were performed with multiple datasets to increase the fit hyper-parameter space. Nb-foil was used as reference, shown in **Fig.S1**. The fitted amplitude value from Nb-foil EXAFS spectra was further used in sample refinements ($S_0^2$=0.99). The fitted values for Nb-foil are reported in **Table S2**.

**2.4. Hydrogenation Properties**

An in-house built Sievert's type apparatus was used for $H_2$ activation, absorption, and desorption from the samples. The detailed description of the instrument is provided elsewhere **(56)**.

All samples were prior activated (50±2 ºC, 40±2 bar) for 17.5 hours following by $H_2$ desorption under dynamic vacuum. The samples were then subjected to one absorption / desorption cycle (50±2 ºC, 40±2 bar, 140 mins), after which the samples were characterised with SR-PXRD and XAS. In a separate experiment, two absorption/desorption cycles were run on TiFe, $Nb_{0.1}$. The experimental error on the $H_2$ content measured in wt% (mass of $H_2$ by mass of metal) was less than 0.05 wt%.

Kinetic analysis was done by linear regression fitting of common diffusion models: a) chemisorption ($\alpha$-model), b) Johnson-Mehl-Avrami nucleation-growth-impingement models in 2 and 3 dimension, respectively (JMA2D, JMA3D), c) contracting volume models in 2 and 3-D, respectively (CV2D, CV3D) f) Ginstling-Brounshtein models 2 and 3-D, respectively (GB2D, GB3D) to understand dominant $H_2$ diffusion mechanisms and related phase transitions **(37-40)**.

## 3. Results and Discussions

### 3.1. Structural characterization with synchrotron powder X-ray diffraction

All the arc-melted samples resulted in the formation of TiFe main phase as reviled by the Rietveld analysis of the SR-PXRD patterns (Figures 1 and 2). Diffraction peaks of pure elemental phases of Fe, or Nb were not observed. Obtained unit cell parameters of TiFe_BH, was similar to the previously reported values **(31) (Table 2)**. The unit cell volume of the Nb-doped samples increased by 1.5% compared to the pure TiFe phases (Table 2) disregarding the Nb content.

In TiFe$_{0.9}$ and Nb-doped samples, the formation of Ti-rich secondary phase $β$-Ti was observed: 2.7 wt%, 3.4 wt%, and 6.4 wt% for the 0, 0.05, and 0.1 content of Nb, respectively. The unit cell volume of the $β$-Ti has also slightly increased in the Nb-doped samples compared to those of pure TiFe$_{0.9}$: by ca. 0.5 and 0.65 Å$^3$ in 0.05 and 0.1 Nb content, respectively. This suggests that increasing the amount of Nb (at(%)) in the starting elemental mixture does not necessarily correspond to a monotonic increase of Nb in the main TiFe phase. On the contrary, an increase in the volume of the secondary $β$-Ti phase (*Im3m*, No. 229) might indicate that this phase is preferred for Nb.

SR-PXRD patterns after the activation, followed by $H_2$ absorption / desorption are shown in **Fig.2 (b), (d)** and **(f)**, respectively. Structural transformation of the secondary $β$-Ti to a somewhat distorted $α$-Ti phase with the space group $P6_3/mmc$ (No. 194) was observed in all the samples with the initial $β$-Ti (top-right insets in **Fig.2 (b), (d), (f)**). The potential nature of the distortion effect on $α$-Ti crystallographic structure is elucidated further with EXAFS analysis (**Section 3.2**).

**Table 2.** Crystallographic data for TiFe_BH, TiFe_AH, TiFe$_{0.9}$_BH, TiFe$_{0.9}$_AH, Nb$_{0.05}$_BH, Nb$_{0.05}$_AH, Nb$_{0.1}$_BH and Nb$_{0.1}$_AH as determined by Rietveld refinement of SR-PXRD data.

| Alloy | Phases (wt %) | Space groups | Lattice parameters (Å) | Cell volume (Å$^3$) | R$_{WP}$ |
|---|---|---|---|---|---|
| **TiFe_BH** | TiFe [100(0)] | *Pm-3m* | a = 2.9870(8) | 26.6510(20) | 0.0792 |
| **TiFe_AH** | TiFe [100(0)] | *Pm-3m* | a = 2.9869(4) | 26.6479(10) | 0.1192 |
| **TiFe$_{0.9}$_BH** | TiFe [97.32(5)] | *Pm-3m* | a = 2.9860(5) | 26.6244(12) | 0.1249 |
| | $β$-Ti [2.68(5)] | *Im-3m* | a = 3.2124(8) | 33.149(2) | |
| **TiFe$_{0.9}$_AH** | TiFe [97.09(23)] | *Pm-3m* | a = 2.98385(7) | 26.566(2) | 0.1361 |
| | distorted_$α$-Ti [2.91(23)] | $P6_3/mmc$ | a = 2.920(3) c = 4.531(3) | 33.47(4) | |
| **Nb$_{0.05}$_BH** | TiFe [96.63(9)] | *Pm-3m* | a = 3.0014(4) | 27.0369(10) | 0.0848 |
| | $β$-Ti [3.37(9)] | *Im-3m* | a = 3.2280(6) | 33.636(2) | |
| Nb$_{0.05}$_AH | TiFe [96.71(8)] | Pm-3m | a = 2.9916(6) | 26.773(2) | 0.1124 |
| | distorted_α-Ti [3.29(8)] | P6$_3$/mmc | a = 2.9472(13) c = 4.658(3) | 35.04(3) | |
| Nb$_{0.1}$_BH | TiFe [93.58(5)] | Pm-3m | a = 2.9987(4) | 26.9650(11) | 0.0871 |
| | β-Ti [6.42(5)] | Im-3m | a = 3.2329(3) | 33.7910(9) | |
| Nb$_{0.1}$_AH | TiFe [92.01(14)] | Pm-3m | a = 2.9921(8) | 26.788(2) | 0.1437 |
| | distorted_α-Ti [7.99(14)] | P6$_3$/mmc | a = 2.9630(10) c = 4.444(2) | 33.783 (2) | |

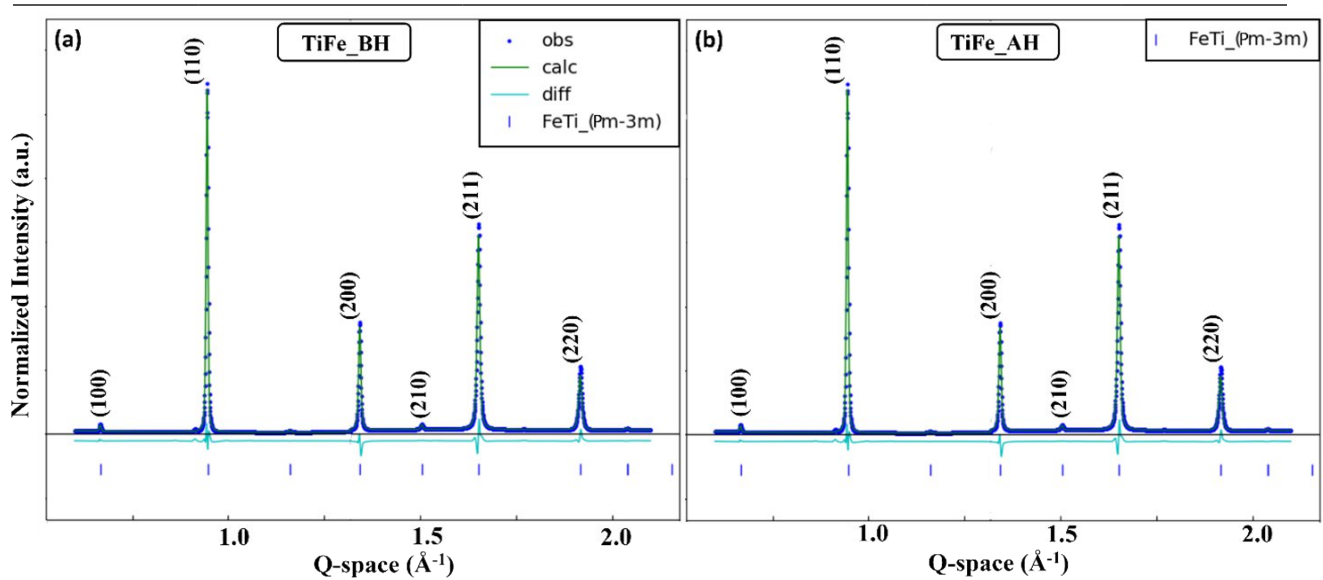

**Fig. 1**. SR-PXRD of arc-melted pristine TiFe metal-alloy systems before H$_2$ exposure: (a) and after activation followed by one H$_2$ absorption/desorption isotherm: (b). Experimental (blue), calculated (green) patterns and their differences (cyan) are shown. The position of the diffraction patterns for each phase is given by the bars below the patterns.

The refinement of Nb occupancies was not performed due to the following reasons: (1) Nb incorporation in both main and secondary phases led to challenges with application of constraints, (2) low wt (%) of the secondary phases would result in unreliable fitting results. Thus, the formation of Ti-rich secondary phases and post-H$_2$ exposure transformation from β-Ti to somewhat distorted α-Ti like structure, were used as predictors during EXAFS analysis. This suggests displacement of Ti/Nb atoms from Ti lattice site of TiFe matrix, as distorted α-Ti can potentially accommodate more Ti atoms in comparison to β-Ti. The above-mentioned point can also be strengthened from variations outside experimental uncertainties, seen in formed secondary phases (wt %) for Nb$_{0.1}$_BH and AH samples, provided in Table 2. As the samples were handled in air, it was important to check for the formation of the common oxide phases TiO$_2$, TiFeO$_3$, Fe$_2$TiO$_x$, FeTiO$_x$ **(24, 57)**. The PXRD patterns have revealed no reflections from the oxides suggesting formation of no or thin oxide surface layers **(58)**.

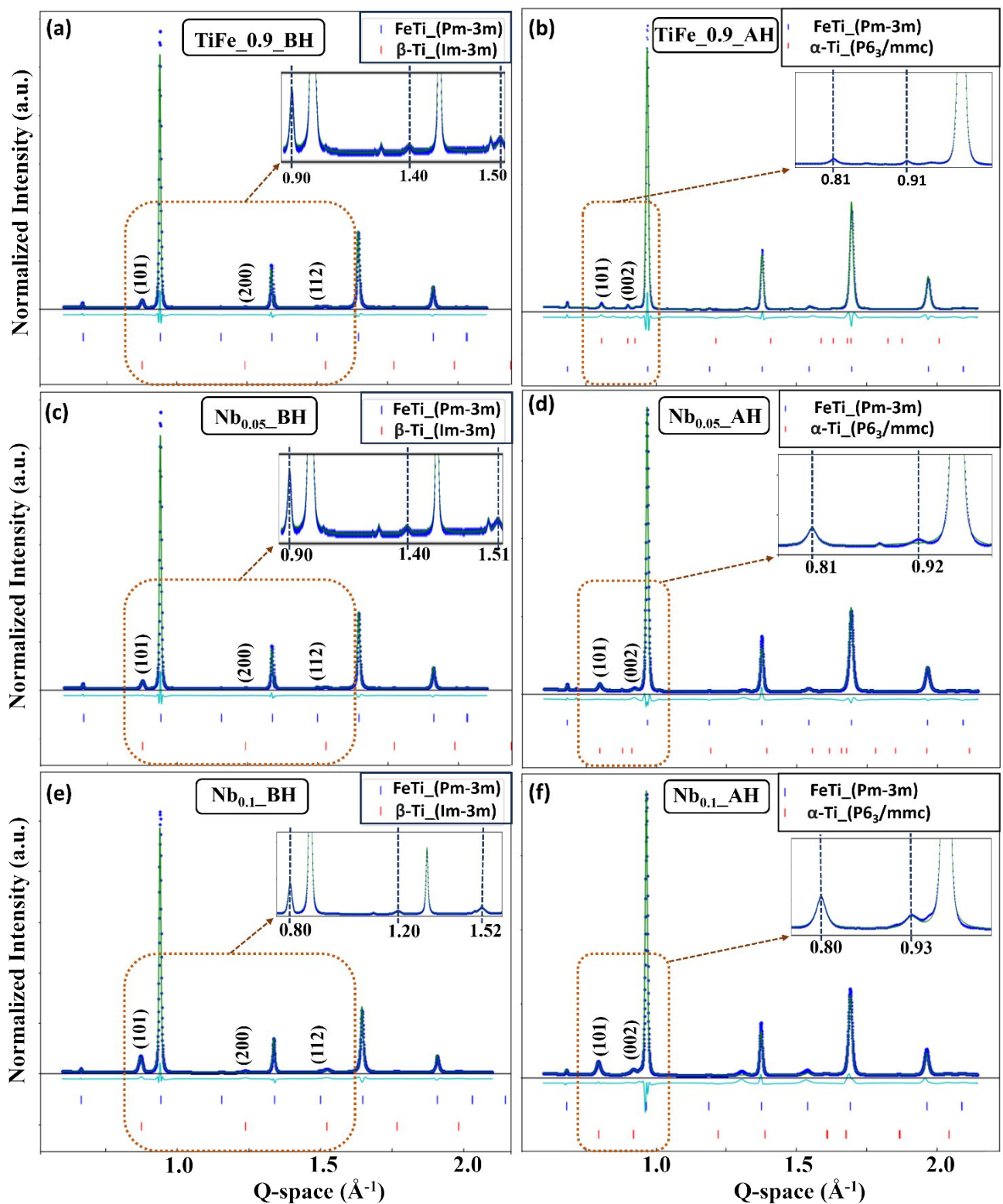

**Fig. 2.** SR-PXRD of arc-melted metal alloy systems before H$_2$ exposure: (a), (c), (e), and after activation followed by one H$_2$ absorption/desorption isotherm: (b), (d), (f). The position of the diffraction patterns for each phase is given by the bars below the patterns.

## 3.2. Local structure analysis of Nb dopants with X-ray absorption spectroscopy

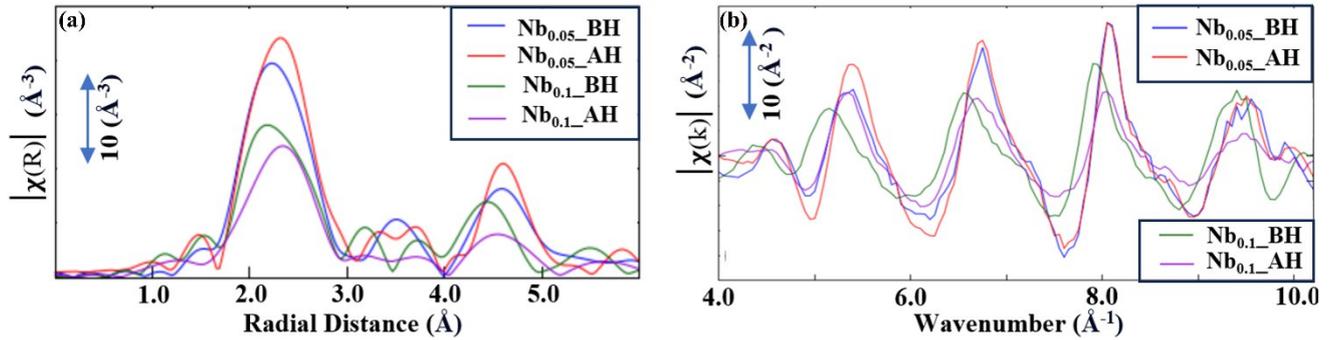

**Fig. 3.** XAS data recorded at Nb k-edge: magnitudes of real and complex imaginary parts of measured EXAFS signals, $k^3$-weighted, phase uncorrected, Fourier transformed (FT) for samples: $Nb_{0.05}\_BH$, $Nb_{0.05}\_AH$, $Nb_{0.1}\_BH$ and $Nb_{0.1}\_AH$, respectively: (a), (b).

X-ray absorption spectroscopy (XAS) is an element-specific, structure-sensitive technique (59) that can elucidate the local coordination environment of Nb and so its preferred location in TiFe and / or secondary phases. Fig. 3 (a) and (b) show the EXAFS spectra of all the Nb-containing samples before and after hydrogen absorption. Nb k-edge absorbance peaks in XAS spectra for BH and AH samples at ~18894.2 eV (Fig. S2) confirm Nb incorporation in all the samples. Furthermore, the differences in the XAS spectra between different samples: $Nb_{0.05}\_BH$, $Nb_{0.1}\_BH$, $Nb_{0.05}\_AH$ and $Nb_{0.1}\_AH$ (Fig. 3(a) and (b)) confirm different Nb environments in the lattices.

To elucidate these environments, we have used various combinations of five model structures to fit the experimental data: two TiFe phases with Nb at Fe (S1) and Ti (S2) sites, Nb in $\beta$-Ti (S3), in $\alpha$-Ti (S4), and Nb in Nb metal (S5) to verify the formation of small Nb clusters not detectable in PXRD. In TiFe, the first coordination shell is made of 8 neighbours at 2.58 Å without considering distance change due to Nb substitution. Fe environment is expected to induce higher peak intensity due to the larger number of electrons. Second coordination shell is composed of 6 neighbours at 2.98 Å, and the third of 12 atoms at 4.21 Å. These scattering paths from the models S1 and S2 were used in the fits.

In $\beta$-Ti phase, the 1st, 2nd and 3rd coordination spheres consist of 8, 6, 12 atoms at 2.84 Å, 3.28 Å, 4.64 Å, respectively. In $\alpha$-Ti phase, the 1st, 2nd and 3rd coordination spheres consist of 2, 12, 6 atoms at 2.83 Å, 2.99 Å, 4.57 Å, respectively. In Nb phase, the 1st, 2nd and 3rd coordination spheres consist of 8, 6, 12 atoms at 2.86 Å, 3.30 Å, 4.67 Å. The phase shift from fitting of Nb foil is about 0.3 Å (Fig. S1, Table S2). Thus, one would expect a contribution from TiFe phase at a lower $R$ (at ca. 2.3 Å considering the phase shift) (Fig. 3a) whereas the contributions from the 1st shells in other structures and the 2nd shell in TiFe are expected at ca. 2.4-2.4 Å. Notably also, only the secondary phases have the scattering intensity between 3.0 and 4.0 Å. The position of the third peak (4-5 Å) would be shifted to lower $R$ in TiFe as well however here the contributions from the double scattering passes are also considerable.

For fitting the EXAFS data, four input supercell (4*4*4) models with Nb distributed homogeneously were created for S1-S4 (Fig. S4 and S6) whereas 1 unit cell was sufficient for the Nb metal environment. The amplitudes ($S_0^2$) of the scattering paths were scaled with the appropriate phase concentrations ($x$ and $y$ for models S2 and S1, respectively) whereas the coordination numbers were set to constant crystallographic values and $S_0^2 = 0.9858$ was obtained from Nb foil fit. The concentration of Nb in the Ti-secondary phases was then obtained as (1 - (x + y)). For BH and AH samples, β-Ti and α-Ti structures (shown in **Fig. 5 (c)**) were used to fit Ti-rich secondary phases, respectively, as determined from the SR-PXRD refinements (**Section 3.1**). The single (SS) and multiple scattering (MS) paths from these coordination geometries were used in fitting $Nb_{0.05}$ and $Nb_{0.1}$ datasets as shown in **Fig. 5 (a)** and **(b)**. Four coordination shells were fitted in radial space (1.25-5.5 Å), which corresponded to k-space (3-10.2 Å$^{-1}$). Changes in the interatomic distances in the cubic phases S1 and S2 were parametrised with one parameter α. To avoid the overparameterization of the EXAFS models, we had to constrain some of the disorder parameters ($\sigma^2$), especially in the higher coordination shells, to constant values (provided in **Table 3**). These values were obtained from multiple fit trials. The spectra of $Nb_{0.05}$ and $Nb_{0.1}$_BH samples were co-refined in one fit that increased the number of independent parameters ($N_{IP}$) to 33, whereas the AH samples were fit one by one hence $N_{IP}$ = 16.

Prior to the final model combinations shown on the Fig. 4, all plausible combinations were fit as summarised in the Supporting information (Fig. S4 – S7), and those with the best results are presented here as the final result. As one can see from the Figures S4 – S7, in the discarded combinations the main disagreement was in the fit of the second coordination peak at ca. 4.5 Å.

The fit results indicate that in the $Nb_{0.05}$ sample the main proportion of Nb is accommodated on the Ti sites in TiFe (ca. 40-60%), and the secondary Ti phases. The amount of Nb in Fe sites is negligible but non-zero within the error bars, and these numbers do not change significantly before and after cycling in hydrogen. In $Nb_{0.1}$, almost half of the initial 40-60 % of Nb in Ti sites of TiFe precipitate into the α-Ti phase, and the number of Nb in Fe sites is small. It can be added here that the fit results without the S1 phase were significantly worse which indicates the non-zero occupancies of Fe sites by Nb in the TiFe.

The obtained occupancy values directly demonstrated that Nb is most likely to occupy Ti lattice sites for all BH and AH samples. With the increasing Nb stoichiometries from 0.05 to 0.1, marginally more Nb was in the β-Ti secondary phase considering the error bars, shown in the bottom-right insets in **Fig, 4(a)** and **(b)**, respectively. This observation also correlates with SR-PXRD patterns of $Nb_{0.05}$ and $Nb_{0.1}$ (discussed in **Section 3.1**), suggesting with increasing Nb content (at%), more Nb incorporation in Ti-rich secondary phases is preferred.

**Table 3.** Data range and metric parameters extracted by fit analysis of the Nb k-edge EXAFS spectra in k-range 3.000 – 10.2 Å$^{-1}$. Amplitude reduction factor $S_0^2$ = 0.9858 was used as obtained from the fitting to reference Nb foil.

| Sample | Fitting range R–Δ (Å) $N_{IP}$ // $N_{FV}$ | Phase fraction (x, y) | Scattering path | R (Å) | N$^c$ | $\sigma^2$ (Å$^2$)$^a$ | ΔE (eV) | R-factor |
|---|---|---|---|---|---|---|---|---|
| $Nb_{0.05}$_BH | 1.3 - 5.2<br>33 // 16 | S1: 0.13(10)<br>S2: 0.55(14)<br>S3: 0.32(17) | S1: <Nb_P0 - Ti_P3><br><Nb_P0 - Fe_P1><br><Nb_P0 - Fe_P2> | 2.573(5)<br>2.971(6)<br>4.202(8) | 8<br>6<br>12 | 0.016(10)<br>0.006(2)<br>0.012$^c$ | 0 | 0.03 |
| | | | S2: <Nb_P0 – Fe_P3><br><Nb_P0 – Ti_P1><br><Nb_P0 - Ti_P2> | 2.573(5)<br>2.971(6)<br>4.202(8) | 8<br>6<br>12 | 0.004(2)<br>0.018(12)<br>0.011(7) | | |
| | | | S3: <Nb_P0 - Ti_P1><br><Nb_P0 - Ti_P2><br><Nb_P0 - Ti_P3> | 2.81(4)<br>3.25(5)<br>4.60(7) | 8<br>6<br>12 | 0.008$^c$<br>0.014$^c$<br>0.011$^c$ | | |
| **Nb$_{0.05}$_AH** | 1.5 - 5.2<br>16 // 8 | S1: 0.12(7)<br>S2: 0.56(10)<br>S4: 0.32(12) | S1: | <Nb_P0 – Ti_P3><br><Nb_P0 – Fe_P1><br><Nb_P0 – Fe_P2> | 2.602(3)<br>3.004(3)<br>4.249(4) | 8<br>6<br>12 | 0.006(1)<br>0.005(3)<br>0.01$^c$ | 0 | 0.01 |
| | | | S2: | <Nb_P0 - Fe_P3><br><Nb_P0 – Ti_P1><br><Nb_P0 - Ti_P2> | 2.602(3)<br>3.004(3)<br>4.249(4) | 8<br>6<br>12 | 0.005(1)<br>0.008(5)<br>0.012(4) | | |
| | | | S4: | <Nb_P0 - Ti_P1><br><Nb_P0 - Ti_P2> | 3.12(4)<br>3.23(5) | 7<br>1 | 0.009$^c$<br>0.009$^c$ | | |
| **Nb$_{0.1}$_BH** | 1.3 - 5.2<br>33 // 16 | S1: 0.05(4)<br>S2: 0.58(8)<br>S3: 0.38(9) | S1: | <Nb_P0 – Ti_P3><br><Nb_P0 – Fe_P1><br><Nb_P0 - Fe_P2> | 2.613(3)<br>3.017(3)<br>4.267(4) | 8<br>6<br>12 | 0.016(1)<br>0.008(1)<br>0.008$^c$ | 0 | 0.03 |
| | | | S2: | <Nb_P0 – Fe_P3><br><Nb_P0 - Ti_P1><br><Nb_P0 - Ti_P2> | 2.613(3)<br>3.017(3)<br>4.267(4) | 8<br>6<br>12 | 0.008(1)<br>0.020(0)<br>0.012(3) | | |
| | | | S3: | <Nb_P0 - Ti_P1><br><Nb_P0 - Ti_P2><br><Nb_P0 - Ti_P3> | 2.98(1)<br>3.45(1)<br>4.86(2) | 8<br>6<br>12 | 0.003$^c$<br>0.01$^c$<br>0.013$^c$ | | |
| **Nb$_{0.1}$_AH** | 1.5 - 5.2<br>16 // 8 | S1: 0.06(5)<br>S2: 0.30(6)<br>S4: 0.64(8) | S1: | <Nb_P0 – Ti_P3><br><Nb_P0 – Fe_P1><br><Nb_P0 - Fe_P2> | 2.621(13)<br>3.027(20)<br>4.280(21) | 8<br>6<br>12 | 0.01(2)<br>0.008$^c$<br>0.015$^c$ | 0 | 0.03 |
| | | | S2: | <Nb_P0 – Fe_P3><br><Nb_P0 -Ti_P1><br><Nb_P0 - Ti_P2> | 2.621(13)<br>3.027(20)<br>4.280(21) | 8<br>6<br>12 | 0.007(3)<br>0.005(2)<br>0.01$^c$ | | |
| | | | S4: | <Nb_P0 - Ti_P1><br><Nb_P0 - Ti_P2> | 3.07(3)<br>3.16(4) | 2<br>2 | 0.013(4)<br>0.017(4) | | |

$N_{IP}$, $N_{FV}$ – number of independent parameters and fitted variables, respectively. $^a$ Debye-Waller factor; $^c$ – set to constant in the fit.

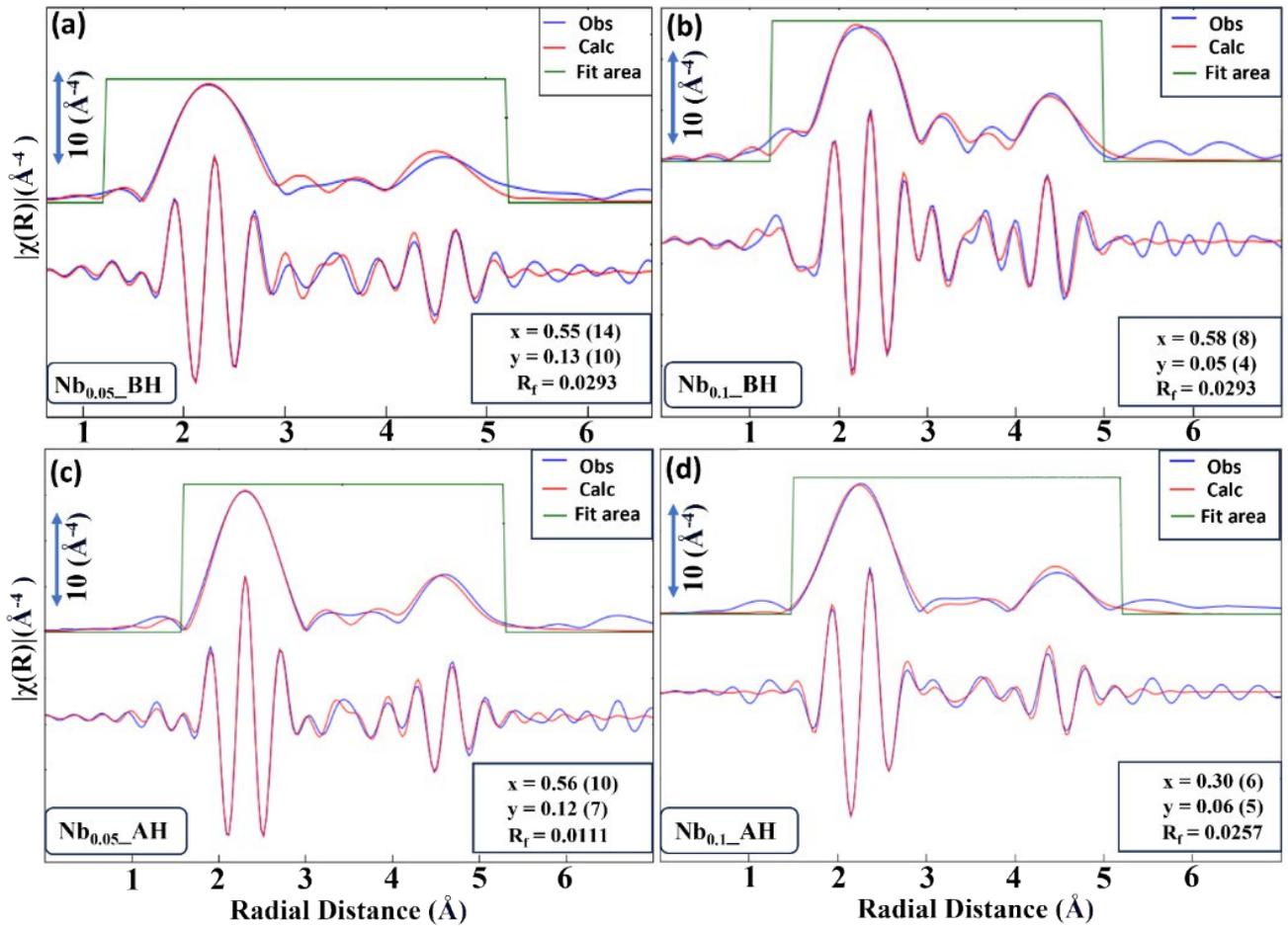

**Fig. 4.** XAS data recorded at Nb k-edge: (a), (b), (c), (d) show Rmr plots with magnitudes (top signals) and real parts (bottom signals) of measured EXAFS signals (blue curves): $k^3$-weighted, phase uncorrected, Fourier transformed (FT) for samples: $Nb_{0.05}\_BH$, $Nb_{0.05}\_AH$, $Nb_{0.1}\_BH$ and $Nb_{0.1}\_AH$, respectively in the K-edge of Nb. Model fits to FT EXAFS functions were performed using Artemis software in the Demeter package (red curves). The fitting regions (green windows) in radial space (~1.25-5.5 Å) correspond to equivalent regions in k-space (3-10.2 Å$^{-1}$).

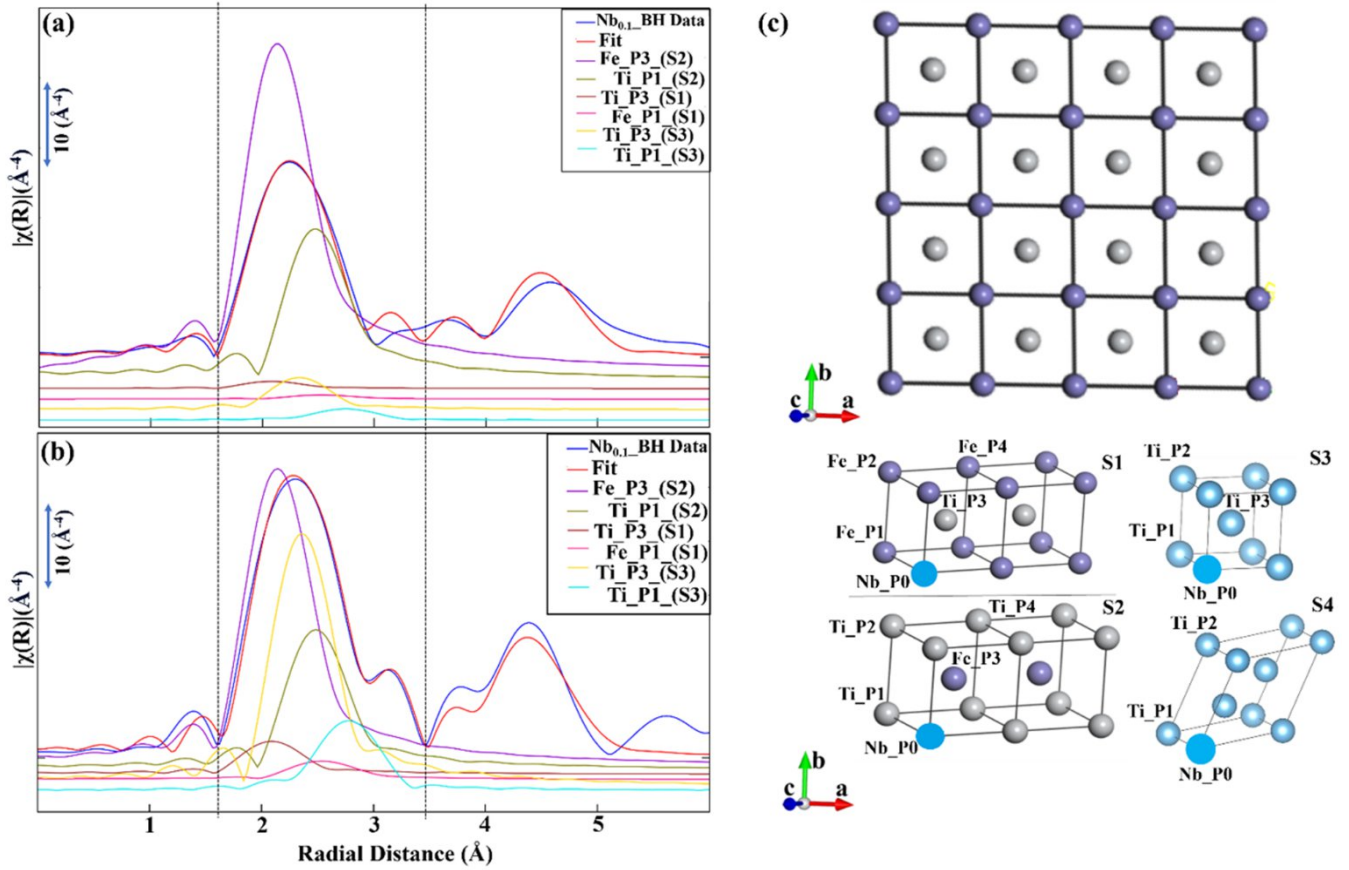

**Fig. 5.** (a) and (b) Proportions of deconvoluted individual scattering path amplitudes and phases from models S1, S2 and S3, respectively used in fitting of experimental FT EXAFS functions for samples: $Nb_{0.05}\_BH$ and $Nb_{0.1}\_BH$. The legend in top-right inset, shows colour coding for individual scattering paths. (c) TiFe supercell (4*4*4): Fe atoms (purple) and Ti atoms (grey). Supercells doped with Nb atom, where the dopant is uniformly distributed in the crystal structure satisfying respective Nb stoichiometries: 0.05, 0.1. For clarity, only single unit cell accommodating one Nb atom are shown. (S1) and (S2) represent models where Nb dopant is placed in Ti and Fe lattice sites, respectively where (S3) and (S4) represent models where Nb is placed in $\beta$ and $\alpha$-Ti-rich secondary phases, respectively. The different atomic positions are represented by P1, P2, P3 and P4 for models S1, S2, S3, and S4, respectively, whereas P0 is the Nb-scatterer position.

**Fig. 5(a), (b)** show the individual deconvoluted scattering path contributions from models S1, S2 and S3 for $Nb_{0.05}\_BH$ and $Nb_{0.1}\_BH$ fitted to EXAFS functions up till 5.5 Å. The scattering path contributions to EXAFS signals for $Nb_{0.05}\_AH$ and $Nb_{0.1}\_AH$ are provided in **Fig. S3(a), (b)**. Differences in path amplitudes and phase contributions originating from model (S3) (Ti_P2_(S3), Ti_P3_(S3)), were seen between samples $Nb_{0.05}\_BH$ and $Nb_{0.1}\_BH$, which showed increased presence of Nb in $\beta$-Ti phase with increasing Nb content. The same effect was observed in the secondary $\alpha$-Ti phase in AH samples. This also corroborates to the increase in the lattice parameter of Ti-rich secondary phases obtained from SR-PXRD refinements. The relatively low magnitudes compared to model (S2), of the contributing paths from model (S1), in both samples $Nb_{0.05}$ and $Nb_{0.1}$, is shown in **Fig. 5(a), (b)** and suggest a low probability of a Nb atom substitution in the Fe lattice site. This is also evident from the fitted fractions $y$ that are essentially zero considering the mathematical errors for all the samples, as shown in **Fig. 4(a-d)**. The change in the Nb coordination environment inferred from the XAS spectra before and after hydrogen sorption can thus be correlated with β→α phase transition in the Ti-secondary phase.

## 4. Hydrogenation Properties

### 4.1. Activation and hydrogen absorption isotherms

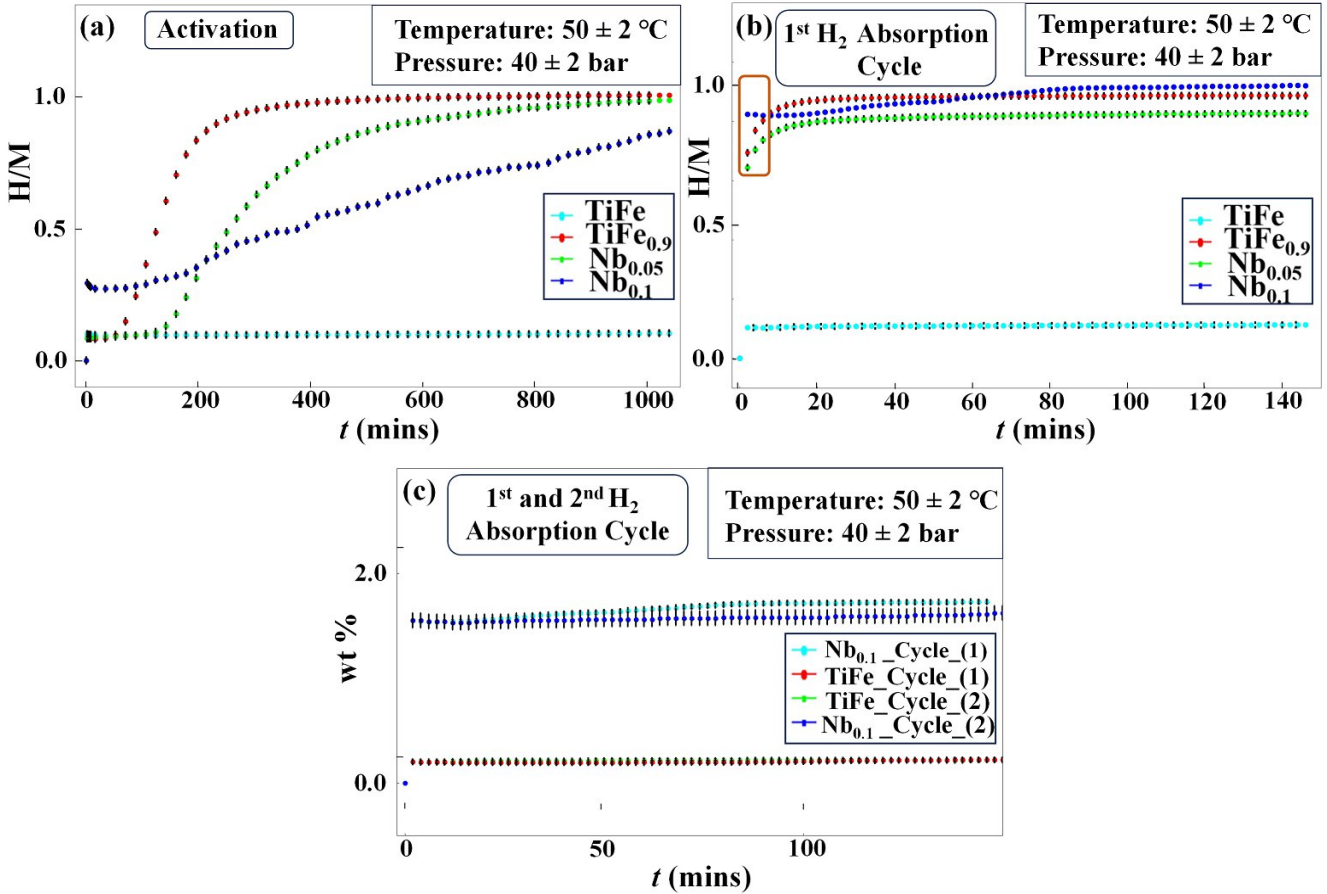

**Fig. 6.** (a) Hydrogen-sorption isotherms during the activation cycle: TiFe (cyan), TiFe$_{0.9}$ (red), Nb$_{0.05}$ (green), Nb$_{0.1}$ (blue) at $T = 50\pm2$ °C and $p(H_2) = 40\pm2$ bar; (b) during the 1$^{st}$ absorption cycle after the activation (at 50±2 °C and 40±2 bar) in H/M ration as function of time (min). Two absorption cycles for samples: TiFe and Nb$_{0.1}$ are shown in (c) to elucidate on reversibility of these alloy systems for H$_2$ (wt %) uptake. The error bars (black) are presented along with the datapoints.

The activation and kinetics isotherms measured at 50±2 °C and 40±2 bar for all the samples are shown in (**Fig. 6(a), (b)**). TiFe$_{0.9}$ and the Nb-doped samples demonstrated significantly improved activation as compared to pristine TiFe (**Fig. 6(a)**). The fast kinetics of H$_2$ absorption during the activation cycle and subsequent cycles in TiFe$_{0.9}$ as compared to TiFe has been explained by the formation of Ti-rich secondary phases **(31)**. Higher Ti content has been known to contribute to enhanced H$_2$ uptake and storage characteristics **(57).** Doping leads to the formation of secondary phases which can have a similar effect on the activation kinetics **(31)**. Herein, we attribute the faster activation of Nb doped samples to two effects occurring at different length scales: (1) at a micrometre scale: the presence of a relatively higher wt (%) of β-Ti secondary phase as compared to TiFe$_{0.9}$ and (2) at an atomic-scale: an increased lattice parameters of the TiFe phase due to the incorporation of Nb in the crystal structure.

A sharp pressure-jump followed by a linear increase was observed for Nb$_{0.1}$ (**Fig. 6(a)**). In contrary, sigmoidal trends in activation isotherms were observed for TiFe$_{0.9}$ and Nb$_{0.05}$. The Nb$_{0.1}$ sample also reached relatively lower H/M ratios, during the activation procedure. Overall, TiFe$_{0.9}$ showed relatively enhanced activation properties in comparison to Nb$_{0.05}$ and Nb$_{0.1}$. The above-mentioned effects might be attributed to a delay in the initial α→β phase transition (H/M range: 0.2-0.6) **(7)**, which requires further investigation. Nonetheless, samples TiFe$_{0.9}$, Nb$_{0.05}$ and Nb$_{0.1}$ showed improved hydrogen absorption during the activation cycle as compared to TiFe.

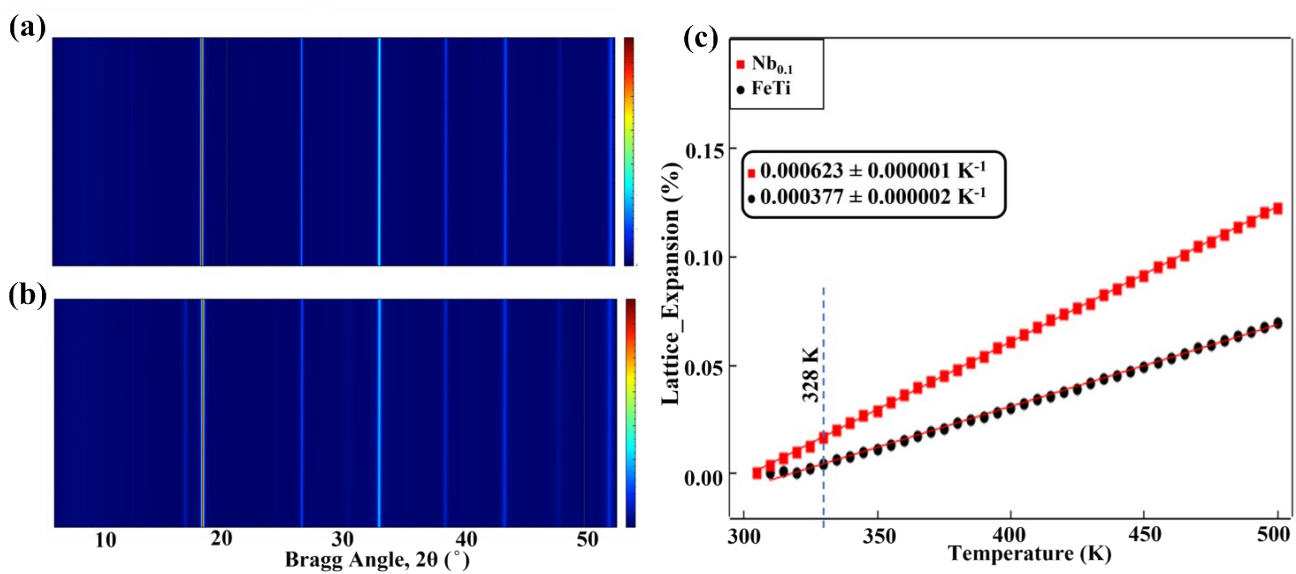

**Fig. 7.** (a), (b) Contour plots for diffraction patterns of samples: TiFe and Nb$_{0.1}$, respectively as function of temperature (300-500 K). Fine shifts in contour lines (shown in plot) to lower Bragg angle 2θ (°) suggests lattice expansion in the alloy systems. (c) Comparison of lattice expansion (%) as function of temperature (K) for the above-mentioned two alloy systems. Results of linear regression fitting is shown in the inset. The dashed blue line at 328 K marks the temperature at which H$_2$ isotherms were recorded.

Additionally, TiFe$_{0.9}$, Nb$_{0.05}$, Nb$_{0.1}$ reached H/M ~ 0.9 in 60 min **(Fig. 6(b))**. The fastest H$_2$ uptake was observed for Nb$_{0.1}$ (**Fig. 6(b)**) reaching 90% of saturation pressure within 3 min. The time ranges for the reactions to reach 90% of its saturation pressures` ($t_{90}$) for the non-equiatomic samples are shown in **Table 4**. This can be attributed to the secondary phase formation and increased lattice parameters due to Nb incorporation into the main TiFe phase. Comparable H$_2$ wt% were observed for TiFe$_{0.9}$ and Nb$_{0.1}$ reaching ~ 1.73.

**Table 4.** Hydrogen adsorption parameters for the pristine and doped samples.

| Samples | $t_{90}$ (min) (range) | H$_2$ (H/M ratio) | H$_2$ (wt%) |
|---|---|---|---|
| TiFe$_{0.9}$_BH | 5.5 – 6.0 | 0.887(9) | 1.73(2) |
| Nb$_{0.05}$_BH | 6.5 – 7.0 | 0.826(8) | 1.58(2) |
| Nb$_{0.1}$_BH | 2.5 – 3.0 | 0.918(9) | 1.72(2) |

The gravimetric capacity (wt%) with Nb incorporation did not show significant decrease as could be expected from the substitution with a heavier element. This could be explained by the larger lattice size that accommodate more H atoms that is also seen from a slight increase in the H/M ratio for $Nb_{0.1}$ sample (**Fig. 6(b)**). Further, the lattice expansion (%) of $Nb_{0.1}$ at 50 ℃ (close to the $H_2$ absorption temperature used in this study), was found significantly greater than that in the pristine TiFe (shown as dotted blue line on **Fig. 7(c)**). The data were obtained from the sequential refinement of the in-situ SR-PXRD patterns as a function of temperature (300-500 K) in vacuum. The lattice expansion rate (%), which was calculated to be almost double for $Nb_{0.1}$ in comparison to TiFe, shown in (**Fig. 7(c)**) and is correlated with improved kinetics for the former system which is discussed in the following section.

$Nb_{0.05}$ showed little to no improvement in $H_2$ uptake and storage properties as compared to $TiFe_{0.9}$. Rather a slight decrease in overall capacity was observed (**Fig. 6(b) and (c)**). This is explained by two concurrent effects: (1) insufficient increase in the lattice parameter with such a low doping amount (**Section 3.1.**) and (2) higher atomic mass of Nb, which had an inverse effect on overall capacity.

To follow the evolution of the secondary phases and its related effects on $H_2$ storage properties, a second isotherm was measured for $Nb_{0.1}$ and TiFe (**Fig. 6(c)**). A slight decrease in $H_2$ wt% was observed for $Nb_{0.1}$ after the second cycle, which was not seen in case of TiFe. This effect can be attributed to retention of $H_2$ in the secondary phases due to incomplete dehydrogenation at R$T$ **(60)**. Indeed, this effect was not observed in TiFe in the absence of the secondary phases. Little to no change in reaction kinetics was observed with cycling.

SEM snapshots of the BH and AH samples were collected and evidenced little effect of hydrogen sorption on sample microscopic morphology (**Fig.S10**). EDS analysis (**Fig.S8**, **S9**, **Table S3**) showed Ti-rich zones further suggesting the presence of Ti-rich secondary phases in $Nb_{0.05}\_BH$ and $Nb_{0.1}\_BH$.

## 4.2. Kinetics Modelling Study

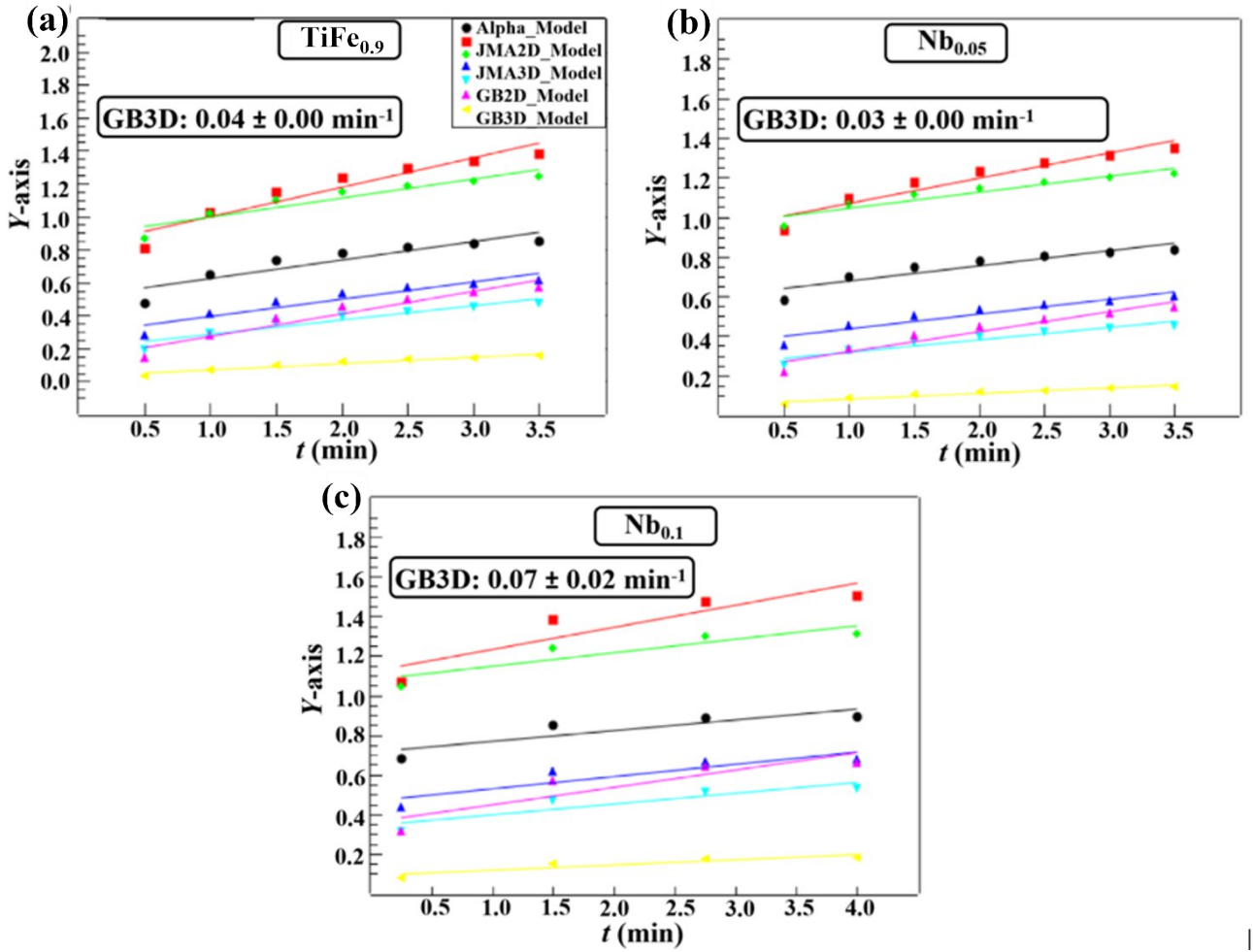

**Fig. 8.** (a), (b), (c) Linear regression fitting of common diffusion models: alpha ($\alpha$), JMA2D, JMA3D, CV2D, CV3D, GB2D and GB3D for samples: TiFe$_{0.9}$, Nb$_{0.05}$ and Nb$_{0.1}$, respectively.
*Left-side of model equations (Table 5) plotted as $Y$-axis.

The hydrogen absorption isotherms (**Fig. 6(b)**) analysed using different kinetics models (**Table 5**) are shown in **Fig. 8(a), (b)** and **(c)**). Reaction rates constants ($k$) obtained from the slopes of the respective linear regression fits, are shown as insets in (**Fig. 8(a), (b)** and **(c)**). Previous studies have reported that the overall hydrogenation reaction kinetics in TiFe is limited by $\alpha \rightarrow \beta$ (TiFeH$_{0.1}$ $\rightarrow$ TiFeH$_{1.04}$) phase transition that occurs in the 0.2 – 0.6 H / M atomic ratios range **(1, 7)**. The linear regression fitting reported in **Fig. 8** was thus carried out in this absorption range.

TiFe$_{0.9}$, Nb$_{0.05}$ and Nb$_{0.1}$ showed best fit with GB3D model (**Table 6**) in agreement with earlier studies **(61)**. Nb incorporation in TiFe crystal structure did not affect the absorption rate-limiting steps despite the significant lattice expansion linked to faster reaction kinetics.

**Table 5.** Rate-limiting-step kinetics models. *Left-side of model equations plotted as *Y*-axis.

| Model Name | Model equations where $\alpha$ = ($\%H_{abs}/\%H_{max}$) | Model Description |
|---|---|---|
| Chemisorption | *$\alpha = kt$ | Surface controlled |
| Nucleation-growth-impingement model (JMA2D) | *$[-\ln(1-\alpha)]^{1/2} = kt$ | 2D growth of existing nuclei with constant interface velocity. |
| Nucleation-growth-impingement model (JMA3D) | *$[-\ln(1-\alpha)]^{1/3} = kt$ | 3D growth of existing nuclei with constant interface velocity. |
| Contracting volume model (CV2D) | *$[1-(1-\alpha)]^{1/2} = kt$ | 2D growth with constant interface velocity. |
| Contracting volume model (CV3D) | *$[1-(1-\alpha)]^{1/3} = kt$ | 3D growth with constant interface velocity. |
| Ginstling-Brounshtien model (GB2D) | *$[(1-\alpha)\ln(1-\alpha)] + \alpha = kt$ | 2D growth, diffusion controlled with decreasing interface velocity. |
| Ginstling-Brounshtien model (GB3D) | *$[1 - (2\alpha/3) - (1-\alpha)]^{2/3} = kt$ | 3D growth, diffusion controlled with decreasing interface velocity. |

**Table 6.** Adjusted $R^2$ values for all model equations for BH samples.

| Samples | Chemisorption | JMA2D | JMA3D | CV2D | CV3D | GB2D | GB3D |
|---|---|---|---|---|---|---|---|
| TiFe$_{0.9}$_BH | 0.754 | 0.858 | 0.835 | 0.844 | 0.869 | 0.891 | 0.914 |
| Nb$_{0.05}$_BH | 0.789 | 0.874 | 0.858 | 0.860 | 0.877 | 0.891 | 0.923 |
| Nb$_{0.1}$_BH | 0.810 | 0.804 | 0.727 | 0.855 | 0.870 | 0.895 | 0.928 |

## 5. Conclusion

In this study, the effects of doping TiFe alloys with small additions of Nb were investigated with respect to their hydrogen sorption properties. The location of Nb in the TiFe crystal structure was characterized by combining the refinement of SR-PXRD and EXAFS data. Strong correlations with the structural properties and the hydrogenation of the metal-alloy systems were established. The detailed findings from this work are summarized as follows.

Non-equiatomic TiFe stoichiometries promoted the formation of secondary β-Ti phase, significantly contributing to hydrogen sorption activation and kinetics. Increasing the Nb content led to the increased β-Ti phase fraction formation. SR-PXRD data showed no peaks due to metallic Nb evidencing its incorporation into the formed phases. TiFe unit cell parameter had increased supporting Nb substitution into this structure. On the other hand, a non-linear dependence between the cell parameter and Nb content also suggested Nb incorporation into the secondary Ti-rich phases. Due to their low quantity, it was not possible to obtain the definite conclusion from refinement of the PXRD data alone. However, further evidence was provided by the XAS study. The refinement of EXAFS spectra showed quantitatively that Nb mostly preferred Ti sites with significant proportion being present in β- and α-Ti phases. To the best of our knowledge, this is the first work that quantitatively reports Nb occupancies in TiFe matrix and its related Ti-rich secondary phases, respectively with EXAFS analysis.

Doping with Nb significantly improves $H_2$ activation and uptake kinetics. Hydrogen absorption data for non-equiatomic compositions: TiFe$_{0.9}$, Nb$_{0.05}$ and Nb$_{0.1}$ showed best fit with GB3D model, suggesting that hydrogen absorption is limited by diffusion. The obtained reaction rate constants were roughly twice for Nb$_{0.1}$ as compared to pristine TiFe. This can be strengthened with results from in-situ SR-PXRD, which show a larger lattice expansion for the Nb$_{0.1}$ sample at 328 K (the temperature of $H_2$ absorption experiment). After $H_2$ abs/desorption, a secondary phase transformation from $\beta$-Ti to a distorted $\alpha$-Ti and related crystallographic details, were reported, utilizing combined SR-PXRD and EXAFS refinements.

In summary, this study shows a considerable improvement of hydrogen activation and kinetics in TiFe alloys with small additions of Nb. These effects were correlated with fundamental understanding of Nb occupancy in crystallographic structure of TiFe. Future investigations are necessary to understand if and how Nb affects $H_2$ cycling behaviours of TiFe-based alloys and how its content can be further optimized for these alloy systems.

## Supporting Info

Supporting information contains further experimental details, details of EXAFS refinement with reference Nb foil and the samples, SEM/EDS data on elemental analysis of BH and AH samples.

## Acknowledgements

A.B. acknowledges the financial support from the Equinor Academia Programme at the University of Stavanger. The authors would like to thank (i) the staff of ESRF (Grenoble, France) beamlines: Dr. D. Chernysov at BM01 (Swiss Norwegian Beamline (SNBL), and BM31 - Dr. Stoian D. for assisting with the in-situ SR-PXRD and EXAFS sample preparation and data collection, respectively. This paper includes data measured within the proposals A31-1-173 and A31-1-209 at BM31, and A01-2-1271 at BM01. The BM31 setup was funded by the Swiss National Science Foundation (grant 206021_189629) and the Research Council of Norway (grant 296087). We also thank Assoc Prof. Diana Lucia Quintero Castro of University of Stavanger (UiS) for sharing valuable beamtime (A01-2-1271). Lastly, we thank John Senith Ravishan Fernando (UiS) for assistance with SEM measurements.

## Declaration of generative AI and AI-assisted technologies in the manuscript preparation process

During the preparation of this work the AI-assisted technologies were not used.